\documentclass[12pt]{article}

\def\tr{{\rm tr}}

\begin{document}

\title{Accelerated cosmological expansion due to a scalar field whose 
potential has a positive lower bound}

\author{Alan D. Rendall\\
Max-Planck-Institut f\"ur Gravitationsphysik\\ Am M\"uhlenberg 1\\
D-14476 Golm, Germany}

\date{}

\maketitle

\begin{abstract}
In many cases a nonlinear scalar field with potential $V$ can lead to
accelerated expansion in cosmological models. This paper contains 
mathematical results on this subject for homogeneous spacetimes. It
is shown that, under the assumption that $V$ has a strictly positive 
minimum, Wald's theorem on spacetimes with positive cosmological 
constant can be generalized to a wide class of potentials. In some cases 
detailed information on late-time asymptotics is obtained. Results on the
behaviour in the past time direction are also presented.  
\end{abstract}

\section{Introduction}

Cosmological models with accelerated expansion are currently of great
astrophysical interest. Two main themes are inflation, which concerns
the very early universe, and the accelerated cosmological expansion 
at the present epoch as evidenced, for instance, by supernova 
observations. The simplest way of obtaining accelerated expansion 
within general relativity is a positive cosmological constant.
A more sophisticated variant is the presence of a nonlinear scalar 
field. For the two themes mentioned above this scalar field is known
as the inflaton and quintessence, respectively. A cosmological constant 
alone may not be an alternative for describing the real universe if the 
observational data indicate different amounts of acceleration at different 
times. Evidence of this kind based on supernova data is presented in 
\cite{alam}. There is at present no good physical understanding of the
exact nature of the inflaton or quintessence and it may even be that both
are the same field, producing different phenomena at different times.

Given the limited information available on the nature of the scalar 
field $\phi$ supposed to cause accelerated expansion, it makes sense 
to study the dynamical behaviour of spacetimes containing a nonlinear
scalar field with a potential taken from a general class satisfying 
a few basic assumptions. The aim of this paper is to do this in 
the context of proving rigorous mathematical theorems which include
anisotropic models. Although the literature on cosmological models 
with accelerated expansion is vast, the number of papers which are 
mathematical in the sense just mentioned, and which go beyond the 
isotropic case, is small. Some which are known to the author are
\cite{wald}, \cite{lee03}, \cite{tchapnda03a}, \cite{tchapnda03b}
and \cite{rendall03a} on the case of a cosmological constant, \cite{moss}
on the case of a quadratic potential (massive scalar field, chaotic
inflation) and \cite{kitada92} and \cite{kitada93} on the case of an 
exponential potential (power-law inflation).

One typical assumption on the scalar field potential $V(\phi)$ is that
it should be non-negative. This is a kind of positive energy assumption
on the field which is very natural. It implies that the dominant 
energy condition is satisfied. The weak energy condition then follows.
Scalar fields with negative potentials have been considered and lead
to interesting dynamical behaviour \cite{erickson}. In the following,
however, it will always be assumed that $V(\phi)\ge 0$. Note that the 
strong energy condition is in general violated even if $V$ is positive 
and that this is intimately connected with the occurrence of accelerated 
expansion.

Consider now spacetimes which are homogeneous but not necessarily 
isotropic. A classical result of Wald \cite{wald} on spacetimes with
positive cosmological constant concerns Bianchi types I-VIII. The
remaining types of spatially homogeneous spacetimes, namely Bianchi
type IX and Kantowski-Sachs, are known to display much more complicated
dynamical behaviour. For this reason consideration will be mainly restricted 
to Bianchi types I-VIII in what follows. A key assumption to be made
on the potential is that it has a strictly positive lower bound. This
is an essential restriction and rules out many of the cases most 
frequently considered in the literature. It is to be considered as the
first step in a more extended investigation and it is chosen as a
starting point because it is likely to be one of the easiest general
classes of potentials to analyse. Nevertheless it will be seen that it
is not trivial to treat rigorously.

To see the meaning of the condition that $V$ is bounded below by a constant 
$V_0>0$ it is useful to consider the following mechanical analogy, which is 
discussed for instance in chapter 11 of \cite{peacock}. The 
equation of motion of the scalar field is
\begin{equation}\label{evscalar}
\ddot\phi+3H\dot\phi+V'(\phi)=0\label{evphi}
\end{equation} 
where $H=-\tr k/3$, $\tr k$ is the mean curvature of the hypersurfaces of 
constant time and a dot denotes $d/dt$. $H$ is positive for an expanding 
cosmological model. This is the equation of 
motion of a ball rolling on the graph of the potential $V$ with a 
variable friction term. The intuitive picture of the motion of the ball
is that it rolls down the potential until it reaches a local minimum where 
it may then oscillate. If the minimum is strictly positive then there are
no oscillations, the phase of accelerated expansion never ends, and the
asymptotic dynamics becomes particularly simple. It is shown in 
section \ref{future} that with an additional mild assumption on the 
behaviour of the potential where it is large a number of the results of 
\cite{wald} can be generalized to the scalar field case. A further 
assumption allows detailed late-time asymptotic properties to be derived
in section \ref{detail}, assuming global existence of solutions in the future.
Section \ref{matter} contains results on global existence of solutions and 
asymptotic behaviour of matter quantities when matter is modelled as
a fluid or collisionless gas.

It is also interesting to know what happens in the past time direction
and this is investigated in Section \ref{past}.
Solutions which exist for infinite proper time in the past often play
the role of inflationary attractors. In the case of a positive cosmological
constant this leads to the de Sitter solution. It is shown that in the 
case of a scalar field with assumptions as before a solution which exists 
for infinite proper time in the past and for which the scalar field remains 
bounded has very special properties. The results for both the past and future 
directions are consistent with the intuition provided by the picture of 
\lq rolling\rq\ and this picture also guided the development of the proofs.

\section{Basics}\label{basics}

Consider a spacetime with vanishing cosmological constant which contains
a nonlinear scalar field and some other matter. The latter is supposed to 
satisfy the strong energy condition and therefore to produce no accelerated 
expansion in the absence of the scalar field. This matter should be thought 
of physically as including baryonic matter, radiation and non-baryonic dark 
matter. The scalar field represents dark energy. The energy-momentum tensor 
of the spacetime being considered is
\begin{equation}
T_{\alpha\beta}=T^M_{\alpha\beta}+\nabla_\alpha\phi\nabla_\beta\phi
-\left[\frac12\nabla^\gamma\phi\nabla_\gamma\phi+V(\phi)\right]
g_{\alpha\beta}.
\end{equation}
where $T^M_{\alpha\beta}$ is the energy-momentum tensor of the matter other
than the scalar field. It is supposed to satisfy the conditions that
for any future-pointing timelike vector fields $X^\alpha$ and 
$Y^\beta$ the inequalities $T^M_{\alpha\beta}X^\alpha Y^\beta\ge 0$  
(dominant energy condition) and 
$[T^M_{\alpha\beta}-(1/2)\tr T^M g_{\alpha\beta}]X^\alpha X^\beta\ge 0$
(strong energy condition) are satisfied. The potential $V$ is assumed to be 
non-negative and sufficiently smooth. For all results in this paper 
$C^2$ is enough. Since $T^M_{\alpha\beta}$ is assumed divergence-free the 
Bianchi identity implies the equation of motion
\begin{equation}
\nabla_\alpha\nabla^\alpha\phi=V'(\phi)
\end{equation}
for the scalar field. 

Now consideration will be restricted to the case that the spacetime is
spatially homogeneous. When expressed in terms of a Gaussian time 
coordinate based on a hypersurface of homogeneity and a left-invariant, 
time-independent frame, the field equations imply the following relations:
\begin{eqnarray}
\frac{dH}{dt}&=&-H^2-\frac{8\pi}3 [\dot\phi^2-V(\phi)]
-\frac13\sigma_{ab}\sigma^{ab}-\frac{4\pi}3(\rho^M+\tr S^M)\label{evtrk}  \\
\frac{d}{dt}(\sigma^a{}_b)&=&-3H\sigma^a{}_b+\tilde R^a{}_b
-8\pi(\tilde S^M)^a{}_b\label{evshear}  \\
H^2&=&\frac{4\pi}3[\dot\phi^2+2V(\phi)]+\frac16(\sigma_{ab}\sigma^{ab}-R)
+\frac{8\pi}3\rho^M\label{ham}
\end{eqnarray}
Here $\sigma_{ab}$ and $\tilde R^a{}_b$ are the tracefree parts of the second 
fundamental form $k_{ab}$ and Ricci tensor of the spatial metric respectively 
and $R$ is the scalar curvature of the spatial metric. The energy density
corresponding to $T^M_{\alpha\beta}$ is denoted by $\rho^M$ while 
$S_{ab}^M$ is the spatial projection of $T^M_{\alpha\beta}$. In the
field equations $S_{ab}^M$ has been split into its trace and tracefree
parts. Combining (\ref{evtrk}) and (\ref{ham}) gives
\begin{equation}
\frac{dH}{dt}=-4\pi \dot\phi^2-\frac12\sigma_{ab}\sigma^{ab}+\frac16 R
-4\pi(\rho^M+\frac13\tr S^M)
\label{evtrk2}
\end{equation}
In the case where the spacetime is a Bianchi model of type I-VIII 
the inequality $R\le 0$ holds \cite{wald}. Assuming that $H$ is positive 
at some time, an assumption which will always be made from now on, it follows 
from (\ref{ham}) that it is positive at all times. For if $H$ vanished
at some time $t_0$ then $\sigma_{ab}$ and $R$ would vanish at $t_0$.
It follows that the induced metric on $t=t_0$ would be flat and the second
fundamental form would vanish. At that time $\rho^M$ would be zero which,
by the dominant energy condition, implies $T^M_{\alpha\beta}=0$. In addition 
$\dot\phi=0$ and $V(\phi)=0$ for $t=t_0$. Now
\begin{equation}\label{energy}
\frac{d}{dt}[\dot\phi^2+2V(\phi)]=-6H\dot\phi^2
\end{equation}
and so in the case under consideration $\dot\phi=0$ and $V(\phi)=0$ at
all future times. {}From (\ref{evscalar}) it follows that $V'(\phi)=0$. The 
unique solution of the field equations with initial data of this type is 
flat spacetime with a constant scalar field and it has $H=0$ at all times, 
contradicting our basic assumption. {}From (\ref{evtrk2}) it can be concluded 
in general that $H$ is non-increasing. 

\section{Late-time dynamics}\label{future}

In this section Wald's theorem \cite{wald} is generalized to the case
of a scalar field whose potential satisfies the following three
assumptions:
\begin{enumerate}
\item $V(\phi)\ge V_0$ for a constant $V_0>0$
\item $V'$ is bounded on any interval on which $V$ is bounded
\item $V'$ tends to a limit, finite or infinite as $\phi$ tends
to $\infty$ or $-\infty$.
\end{enumerate} 
While condition 1. is strong, conditions 2. and 3. are satisfied by many
potentials considered in the literature. Condition 2. is used in 
\cite{miritzis}.

\noindent
{\bf Theorem 1} Consider a solution of the Einstein equations of Bianchi
type I-VIII coupled to a nonlinear scalar field with potential $V$
of class $C^2$ satisfying conditions 1. - 3. above and other matter 
satisfying the dominant and strong energy conditions. If the solution is 
initially expanding ($H>0$) and exists globally to the future then 
for $t\to\infty$, the quantites $\sigma_{ab}\sigma^{ab}$, $R$,
$\rho^M$ and $H^2-(8\pi/3)[\dot\phi^2/2+V(\phi)]$ decay
exponentially. $V(\phi)$ converges to some constant $V_1$, 
$V'(\phi)\to 0$ and $H\to (8\pi V_1/3)^{1/2}$. 

\noindent
{\bf Proof} Since $H$ is bounded in the future it follows from 
(\ref{evtrk2}) that $\int_{t_0}^\infty \dot\phi^2 dt<\infty$ for any 
initial time $t_0$. {}From (\ref{ham}) it follows that $V(\phi)$ is bounded 
to the future. Assumption 2. then implies that $V'(\phi)$ is bounded.
Putting this information into (\ref{evscalar}) and using the
boundedness of $\dot\phi$ shows that $\ddot\phi$ is bounded.
Combining this fact with the integrability of $\dot\phi^2$
shows that $\dot\phi\to 0$ as $t\to\infty$. The quantity
$\dot\phi^2+2V(\phi)$, being monotone and bounded below, tends to a 
limit $2V_1$ as $t\to\infty$. Assumption 1. implies that $V_1>0$.
We have $V(\phi)\to V_1$. Consider now the quantity 
\begin{equation}
Z=9H^2-24\pi[\dot\phi^2/2+V(\phi)]
=\frac32(\sigma_{ab}\sigma^{ab}-R)+24\pi\rho^M
\end{equation}
which is inspired by the quantity $S$ used in \cite{moss}.
\begin{equation}
\frac{dZ}{dt}=-6H\left[\frac32\sigma_{ab}\sigma^{ab}-\frac12 R
+4\pi(3\rho^M+\tr S^M)\right]\le -3HZ
\end{equation}
Since $H\ge (8\pi V_0/3)^{1/2}$ it follows that $Z$ decays exponentially
as $t\to\infty$, implying several of the assertions of the theorem. In
particular the statement on the convergence of $H$ is established.
Let $\phi_2$ and $\phi_3$ be the infimum and supremum respectively
of those numbers arising as limits of a sequence $\{\phi(t_n)\}$ for some
sequence $\{t_n\}$ tending to infinity. If $\phi_2<\phi_3$ then $V'$ must
vanish on the entire interval $(\phi_2,\phi_3)$. It follows that in
this case $V'(\phi)\to 0$ as $t\to\infty$. Otherwise, $\phi_2=\phi_3$
and $\phi$ converges to a limit, finite or infinite. Under the
assumptions of the theorem it then follows that $V'(\phi)$ converges
as $t\to\infty$. If this limit were non-zero then $\ddot\phi$ would 
converge to a non-zero limit, contradicting the fact that 
$\dot\phi\to 0$. Hence in fact $V'(\phi)\to 0$ and this completes the 
proof of the theorem.

The statement of the theorem can now be compared with the \lq rolling\rq\
picture. For this it is useful to suppose that any points where
$V'(\phi)=0$ are isolated. Then the theorem implies that $\phi$
converges as $t\to\infty$ to a finite limit $\phi_1$ with $V'(\phi_1)=0$
or to plus or minus infinity. Thus the scalar field converges to a critical 
point of the potential (possibly at infinity) and the behaviour of $H$ 
can be interpreted as the emergence of an effective cosmological constant 
$V_1$ at late times. The critical point $\phi_1$ does not have to be a 
minimum of the potential and some further remarks on the case it is a maximum 
will be made in Section \ref{past}. Note that in some cases it can be 
predicted which critical point of $V$ the solution converges to. Using
the fact that $V(\phi(t))$ can never exceed the value of 
$\dot\phi^2/2+V(\phi)$ at any time $t_1$ with $t_1<t$ it can be shown
that the solution gets trapped in a local minimum of the potential.

To end this section some comments will be made on Bianchi type IX models.
These are known to include solutions which recollapse. For a
restricted class of initial data the results of Theorem 1 carry over. An
important observation in \cite{wald} is that although the spatial 
scalar curvature may be positive for type IX it satisfies an inequality
of the form $R\le \bar R (\det g)^{-1/3}$ where $\det g$ is the determinant 
of the spatial metric and $\bar R$ is the scalar curvature of the isotropic
metric with unit determinant. Suppose that at some initial time $t_0$ the 
inequality $16\pi V_0-\bar R(\det g)^{-1/3}>0$ holds and that $H$ is 
initially positive. It follows that the determinant of the spatial metric is 
increasing and that the initial inequality is preserved. Thus $H$ remains 
positive. If the solution exists globally in time then $R$ decays 
exponentially and $H$ tends to a limit as $t\to\infty$. As a consequence
$Z$ decays and this gives statements on decay as in Theorem 1. Recall that 
the closed universe recollapse conjecture which was proved in \cite{lin} 
implies that a Bianchi type IX spacetime with matter satisfying the dominant
and strong energy conditions cannot expand for ever. The above discussion,
together with suitable global existence statements, shows that a scalar field 
(which violates the strong energy condition) prevents the analogue of the 
result of \cite{lin} from holding. The arguments leading to global 
existence are described in Section \ref{matter}.

\section{Detailed asymptotics in the future}\label{detail}

In this section the results of Theorem 1 will be refined in order to get more 
detailed information about the asymptotics of the solutions at late times
under additional assumptions on the potential. Suppose that a solution is as 
in Theorem 1 so that $\phi$ tends to a finite limit $\phi_1$
as $t\to\infty$. Without loss of generality it may be assumed by a 
redefinition of the potential that $\phi_1=0$. It follows from Theorem 1
that $V'(0)=0$. In this section it will be assumed in addition that
$V''(0)=\beta>0$ so that $V$ has a non-degenerate minimum at the origin.
Given $\epsilon>0$  the estimates
\begin{eqnarray}
(\beta-\epsilon)\phi^2&\le& V(\phi)-V_1\le(\beta+\epsilon)\phi^2\label{quad1}
  \\
(\beta-\epsilon)\phi^2&\le& \phi V'(\phi)\le(\beta+\epsilon)\phi^2\label{quad2}
\end{eqnarray}
hold for $\phi$ sufficiently small. A disadvantage of (\ref{energy}) is 
that only one of the terms 
differentiated on the left hand side appears on the right hand side. If 
the other term also appeared we could hope to prove a rate of decay for
the energy-like quantity on the left hand side. As it is we can only show 
that this quantity is not increasing, The situation can be improved by 
modifying the energy, adding a term $\alpha\phi\dot\phi$ for a suitable
constant $\alpha$. This device has previously been used in studying solutions
of the Einstein equations in \cite{choquetmoncrief} (section 10) and
\cite{ringstrom} (section 4). 
\begin{eqnarray}
&&d/dt [\dot\phi^2/2+(V(\phi)-V_1)+\alpha\phi\dot\phi]\nonumber \\
&&=(-3H+\alpha)\dot\phi^2-3\alpha H\phi\dot\phi-\alpha\phi V'(\phi)
\end{eqnarray} 
Note that
\begin{equation}
|\phi\dot\phi|\le \frac12\dot\phi^2+\frac1{2(\beta-\epsilon)}(V(\phi)-V_1)
\le C\left(\frac12\dot\phi^2+V(\phi)-V_1\right)
\end{equation}
for a suitable constant $C$. It can be concluded that for $\alpha$ 
sufficiently small
\begin{equation}
\frac12\dot\phi^2+V(\phi)-V_1\le C\left(\frac12\dot\phi^2+V(\phi)-V_1
+\alpha\phi\dot\phi\right)
\end{equation}
For $\alpha$ small enough it can also be concluded that 
\begin{equation}
(-3H+\alpha)\dot\phi^2-3\alpha H\phi\dot\phi-\alpha\phi V'(\phi)
\le -C(\dot\phi^2+V(\phi)-V_1+\alpha\phi\dot\phi)
\end{equation}
for a positive constant $C$. Putting these facts together shows that 
$\dot\phi$ and $V(\phi)-V_1$ decay exponentially for $t\to\infty$.
This in turn implies that the convergence of $H$ to its limit is
exponential. With this information in hand it is possible to proceed
as in \cite{lee03} to show that $g_{ab}=e^{2H_1t}(g^0_{ab}
+O(e^{-\delta H_1t}))$ for some $\delta>0$, $H_1=\sqrt{8\pi V_1/3}$, and
a metric $g^0_{ab}$ not depending on time. The inverse metric has a 
corresponding expansion. In the course of the proof it is also shown
that $\sigma^a{}_b=O(e^{-H_1t})$. The following has been shown:

\noindent
{\bf Theorem 2} Consider a solution of the Einstein equations satisfying
the hypotheses of Theorem 1 for which $\phi\to\phi_1$ as $t\to\infty$.
If $V''(\phi_1)>0$ then as $t\to\infty$ the quantities $\dot\phi$,
$V(\phi)-V(\phi_1)$ and $H-H_1$ decay exponentially as $t\to\infty$
while $e^{-2H_1 t}g_{ab}$ converges to a limit. Here 
$H_1=\sqrt{8\pi V(\phi_1)/3}$.

\section{Perfect fluids and collisionless matter}\label{matter}

All the results above have been based on assuming a solution which exists
globally to the future. To go beyond this it is necessary to introduce a
specific matter model and to prove a global existence theorem for it.
After that the asymptotics of the matter fields can be studied in more
detail. Consider for a moment the case of a cosmological constant instead
of a scalar field. In that case global existence and asymptotic behaviour
have been analysed in \cite{lee03} for collisionless matter described by
the Vlasov equation. For an untilted perfect fluid with a linear equation 
of state and a cosmological constant a dynamical systems analysis of 
models of Bianchi class A has been carried out in \cite{coley}. In the 
following analyses for a nonlinear scalar field will be carried out for 
both perfect fluids with linear equation of state and collisionless matter.

Now the assumptions already made will be specialized to the case where
$T^M_{\alpha\beta}$ is the energy-momentum of a perfect fluid with 
linear equation of state or a collisionless gas. For background on the
latter case see \cite{rendall02}. Before tackling the scalar field 
let us look at the case of a cosmological constant and perfect fluid as
a matter model. Thus 
\begin{equation}
T^M_{\alpha\beta}=(\mu+p)u_\alpha u_\beta+pg_{\alpha\beta}
\end{equation}
and $p=(\gamma-1)\mu$. Assume that $1\le\gamma<2$.
It will be shown that the solutions exist globally to the future and
their asymptotics will be compared with the expansions given by
\cite{starobinsky}. The basic equations for the fluid, which we take 
from \cite{rendall03a}, are
\begin{eqnarray}
\rho&=&\mu(1+\gamma|u|^2)                        \\
j^a&=&\gamma\mu(1+|u|^2)^{1/2}u^a          \\
S^a{}_b&=&\mu[\gamma u^au_b+(\gamma-1)\delta^a_b]
\end{eqnarray}
and
\begin{eqnarray}
\frac{d}{dt}\rho+3H\rho+H\tr S&=&-\nabla_a j^a+\sigma^a{}_b S^b{}_a
\\
\frac{d}{dt}(j^a)+5Hj^a&=&-\nabla^b S^a{}_b+2\sigma^a{}_b j^b
\end{eqnarray}
In order to stop the notation becoming too cluttered the superscript $M$
has been omitted from the matter quantities here.

The Einstein equations with positive cosmological constant coupled to
a perfect fluid constitute a system of ordinary differential equations
(ODE). Standard ODE theory says that a local solution can be extended
as long as $g_{ab}$, $k_{ab}$, $(\det g)^{-1}$, $\rho$, $\rho^{-1}$ and
$u^a$ are bounded. If it can be shown that these quantities are bounded
for any solution on an interval $[t_0,t_1)$ then global existence 
follows. It will now be shown that they can be bounded. The 
procedure used in the proof of Wald's theorem shows that 
$\sigma_{ab}\sigma^{ab}$ and $\tr k$ are bounded. Arguing as in \cite{lee03},
Lemma 1 and 2, then shows that $g_{ab}$, $k_{ab}$ and 
$(\det g)^{-1}$ are bounded. With this information it can be seen that
$|d\rho/dt|$ can be bounded by $C\rho$ for a suitable constant $C$. Hence
$\rho$ and $\rho^{-1}$ are bounded.  Next the evolution equation for
$j^a$ can be used to show that $j^a$ is bounded. Since $u^a$ 
can be expressed as a smooth function of $\rho$ and $j^a$ \cite{rendall03a}
it follows that $u^a$ is bounded. Thus global existence has been proved.

Let $H_1=\sqrt{\Lambda/3}$. Consider now the untilted case where, by 
definition, $u^a=0$. Then $\rho=\mu$ and $\mu$ 
satisfies the equation $\dot\mu=-3\gamma H\mu$. This implies that 
$d/dt(e^{3\gamma H_1t}\mu)$ is equal to an exponentially decaying factor 
times $e^{3\gamma H_1t}\mu$
and it follows that $\mu=\mu_0 e^{-3\gamma H_1t}+o(e^{-3\gamma H_1t})$. In
the tilted case it is useful to first look at the behaviour of $|j|$.
\begin{equation}\label{modj}
\frac{d}{dt}(|j|^2)=-8H|j|^2+2\sigma_{ab}j^aj^b-2j_a\nabla^b S^a{}_b
\end{equation}
Note the inequality $|u^a|\le |u|e^{-H_1t}$ where the expression on the 
left denotes the modulus of the components while that on the right denotes
the length of the vector with respect to the spatial metric. A similar
relation holds for any vector. Now 
\begin{equation}
\nabla^b S^a{}_b=\gamma\nabla_b u^a (\mu u^a)+\gamma\nabla_b u^b (\mu u^a)
\end{equation}
Putting together these facts shows that the last term in (\ref{modj}) can
be bounded by $C|j|^2e^{-H_1t}$. Since the second term on the right hand 
side of (\ref{modj} admits a similar
bound it can be concluded that $|j|=\bar j e^{-4H_1t}+o(e^{-4H_1t})$ for some
constant $\bar j$. Putting this information back into the equation for
$j^a$ shows that $j^a=j^a_0e^{-5H_1t}+o(e^{-5H_1t})$. The comparison with the
expansions of \cite{starobinsky} is as follows. The leading terms in
the expansion of the geometry have been validated, as have those of the 
matter quantities in the untilted case. For a tilted fluid the
leading order behaviour of $j^a$ has been determined but the behaviour of
$\mu$ and $u^a$ remains open.

In a spacetime as in the previous section where the matter content is
a perfect fluid with linear equation of state it is possible to proceed
in a very similar way to what was done for a cosmological constant. Once
again the global existence question involves controlling the solution of
a system of ODE. To ensure continuation of a solution it is necessary
to control the quantities listed previously together with $\phi$ and 
$\dot\phi$. The boundedness of $\dot\phi$ follows from that of $H$
and the boundedness of $\phi$ is an easy consequence. {}From this point
on the arguments used in the case of a cosmological constant apply and
the same statements about asymptotics are obtained.

When matter is described by the Vlasov equation the asymptotics for 
a spacetime with positive cosmological constant have been determined by Lee
\cite{lee03}. For a spacetime with scalar field as in the last section
global existence holds \cite{lee04}. Under the hypotheses of Theorem 2 
we obtain a lot of information about the asymptotic behaviour of the 
solution as $t\to\infty$. The methods of \cite{lee03} allow  
further information to be obtained, which will now be summarized
briefly. The generalized Kasner exponents converge to $1/3$. The
components $V_i$ of the velocity of a particle converge exponentially
to a limit. The spacetime is future geodesically complete. Decay rates
can be obtained for the components of the energy-momentum tensor. For
instance $\rho=O(e^{-3H_1t})$ and $|j|=O(e^{-4H_1t})$. These should be 
compared with the results for dust given above. More details of 
these methods can be found in \cite{lee03}.

\section{Dynamics in the past time direction}\label{past}

A spatially homogeneous spacetime with zero cosmological constant and
matter satisfying the strong energy condition which is expanding at one
time never exists for an infinite proper time towards the past. This 
statement is related to the Hawking singularity theorem and is easy to prove
directly. It follows from the differential inequality $dH/dt\le -H^2$
that if $H$ is positive at some time it must blow up in finite time
towards the past. If we keep the strong energy condition but introduce
a positive cosmological constant there is a solution which exists 
globally towards the past, namely the de Sitter solution with 
spatially flat slicing. Its metric is 
\begin{equation}
-dt^2+e^{2H_1t}(dx^2+dy^2+dz^2)
\end{equation} 
In fact it is the only spacetime with these properties, as will now be
shown. The Hamiltonian constraint implies that $H^2\ge\Lambda/3$. If
equality holds in this relation at some time then, since $H$ is 
non-increasing it must hold at all earlier times. This equality implies
that $\rho^M=0$, so that the solution is vacuum, and that $\sigma_{ab}=0$,
so that it is isotropic. Furthermore $R=0$, so that it is spatially 
flat. Hence the spacetime is de Sitter. The remaining case to be 
considered is that where $H^2>\Lambda/3$ at some time. Then this also
holds at all earlier times. There exists $\epsilon>0$ such that
$H^2(1-\epsilon)>\Lambda/3$ and hence $dH/dt\le -\epsilon H^2$.
As a consequence $H$ blows up at a finite time in the past. 

Although the de Sitter spacetime in the form given above exists for an 
infinite amount of proper time towards the past, it is not past 
geodesically complete. These coordinates only cover half of de Sitter space. 
As $t\to-\infty$ a Cauchy horizon is approached. In the case where 
the hypersurfaces of constant time can be compactified by factoring out
by a discrete group of isometries these results follow from general 
theorems of Andersson and Galloway \cite{galloway} which do not require
any symmetry assumptions. 

In the case of a scalar field the Hamiltonian constraint implies
$H^2\ge (8\pi/3)V(\phi)$. If equality holds at some time $t_0$ then, as 
in the case of a cosmological constant, the spacetime is isotropic and 
spatially flat and the matter other than the scalar field vanishes. In 
addition, $\dot\phi=0$ for $t=t_0$. The unique solution with these data 
is the de Sitter solution with a scalar field which is time independent.
The evolution equation for $\phi$ implies that $V'(\phi)=0$. If $V$
has critical points solutions of this type exist and otherwise there
are no such solutions. The geometry is simply the de Sitter geometry,
with the constant scalar field replacing the cosmological constant.
Let us call these pseudo-de Sitter solutions. Suppose now that 
equality never holds and that the solution exists globally to the past
with the potential remaining bounded. Under these circumstances
several of the steps in the analysis of the expanding direction can be 
repeated.

Let $V_2$ be an upper bound for $V(\phi)$ in the past time direction.
If $H^2$ ever exceeded $(8\pi/3)V_2$ then the argument used in the case
of a cosmological constant would prove the existence of a singularity
after a finite time to the past. Thus under the given assumptions on the 
solution $H$ is globally bounded towards the past. It can be concluded 
successively that $\dot\phi$, $V'(\phi)$ and $\ddot\phi$ are bounded and
that $\dot\phi\to 0$ as $t\to -\infty$. Then $V(\phi)$ converges to a
limit $V_1$. If $V_1$ were less than $3/8\pi$ times the limit of 
$H^2$ then there would have to be a singularity, a contradiction. 
The inequality in the opposite direction follows from the Hamiltonian
constraint. Hence $H^2$ converges to $(8\pi/3)V_1$ as $t\to-\infty$. 
The quantities $\sigma_{ab}\sigma^{ab}$, $R$ and $\rho^M$ converge to
zero. The evolution equation for $\phi$ implies that $V'(\phi)\to 0$.

Do any solutions of the type just discussed actually exist? This can
be answered in the affirmative using a dynamical systems analysis 
due to Foster \cite{foster}. In the case where the potential has a local 
maximum at $\phi_1$ there are solutions for which $\phi\to\phi_1$ as
$t\to -\infty$. This represents an instability of the pseudo-de Sitter
solutions and has been called the deflationary universe by Barrow
\cite{barrow}.

It follows from (\ref{energy}) that the limit $\phi_1$ cannot be a strict 
local minimum of the potential. If the potential is such that all critical
points of $V$ are strict local minima then it can be concluded that for
any solution which exists globally to the past the scalar field cannot be 
bounded at early times.

It would be interesting to have further information about the nature 
of the singularity in the past in the case that $H$ does blow up in
finite time. In the isotropic and spatially flat case it is known that
there are many solutions where the potential becomes negligible near
the singularity and the dynamics resembles that for a massless linear 
scalar field. Whether all finite-time singularities in spatially 
homogeneous solutions of the Einstein equations coupled to a nonlinear
scalar field are of this type is not known.

\section{Conclusions}

Let us sum up what has been learned about scalar fields whose potential
has a positive lower bound. Wald's theorem about the asymptotics of
homogeneous spacetimes with positive cosmological constant which exist 
globally in the future has been generalized to this class of scalar 
fields. This is a cosmic no hair theorem for this setting. It confirms
the intuitive picture where the scalar field rolls down to a minimum
of the potential. Under a non-degeneracy assumption on the minimum these
asymptotics can be refined. For some interesting phenomenological matter
models (perfect fluid with a linear equation of state, collisionless 
matter) solutions exist globally in the future and various aspects of
the asymptotic behaviour of geometry and matter can be determined.
For collisionless matter and untilted perfect fluids the picture 
established is rather complete. For tilted perfect fluids only partial
results were obtained.

As stated in the introduction the present investigation is the first 
step in obtaining a mathematical understanding of the dynamical effects
of nonlinear scalar fields. The next case which it would be natural to
look at is that of power-law and intermediate inflation. This would
mean considering a potential whose lower bound is zero but for which
this lower bound is not attained for any finite value of $\phi$.
For power-law inflation, where $V(\phi)=e^{-\lambda\phi}$, results are 
available in \cite{kitada92} and \cite{kitada93}. The proofs are quite 
intricate and inflationary behaviour only occurs if $\lambda$ is not 
too large. Intuitive considerations suggest that there should be 
generalizations of the results of this paper under the assumption that the
potential does not fall off too fast at infinity. It is to be expected that
the behaviour of the spacetime at late times is monotone and that the
accelerated expansion continues forever.

A more complicated case is that of chaotic inflation. The standard 
potential is $V(\phi)=m^2\phi^2/2$ but similar effects are expected 
for any potential which is zero for some finite value of $\phi$. In
this case the accelerated expansion does not last for ever and at
late times the scalar field oscillates. In order to control the
dynamics it seems that some kind of averaging techniques would be 
necessary. A positive cosmological constant simplifies the late-time
dynamics of solutions of the Einstein equations and a potential with
a positive lower bound does the same. This allows all Bianchi types 
I-VIII to be handled in a unified way. This favourable situation can
be expected to persist for power-law and intermediate inflation. For
chaotic inflation, however, the late-time asymptotics is likely to be 
much more like that found when the cosmological constant vanishes and
there is no nonlinear scalar field. Then the different Bianchi types
have individual behaviour and much less is known. For an example of what
can happen, see \cite{ringstrom2}.

Of course eventually we would like to handle inhomogeneous spacetimes
with as little symmetry as possible. In \cite{tchapnda03a} and 
\cite{tchapnda03b} the asymptotics of plane symmetric solutions
of the Einstein-Vlasov system with positive cosmological constant were
determined. Comparable results for spacetimes with a scalar field are
not available. The asymptotics of spacetimes with positive cosmological
constant which are general (i.e. without symmetries) has been discussed
in \cite{rendall03a}. Presumably a similar discussion in possible for
scalar fields with a positive minimum or for power-law inflation. 
Formal series solutions for the latter case have been written down in
\cite{mueller}. 

In conclusion, it is clear from the above discussion that cosmological 
models with accelerated expansion give rise to many interesting challenges 
for mathematical relativity. The dynamics of these spacetimes should be
examined under general hypotheses and the results compared wherever
possible with the conclusions drawn from the rapidly growing body of
observational data.


\begin{thebibliography}{12}

\bibitem{alam} Alam, U., Sahni, V., Saini, T. D. and Starobinsky, A. A. 2003 
Is there supernova evidence for dark energy metamorphosis? Preprint 
astro-ph/0311364.
\bibitem{galloway} Andersson, L. and Galloway, G. 2002 dS/CFT and spacetime
topology. Adv. Theor. Math. Phys. 6, 307-327.
\bibitem{barrow} Barrow, J. D. 1986 The deflationary universe - an
instability of the de Sitter universe. Phys. Lett. B 180, 335-339.
\bibitem{choquetmoncrief} Choquet-Bruhat, Y. and Moncrief, V. 2001 Future 
global in time einsteinian spacetimes with $U(1)$ isometry group. Preprint
gr-qc/0112049.
\bibitem{coley} Coley, A. A. and Wainwright, J. 1992 Qualitative analysis 
of two-fluid Bianchi cosmologies. Class. Quantum Grav. 9, 651-665.
\bibitem{erickson} Erickson, J. K., Wesley, D. H., Steinhardt, P. J. and 
Turok, N. 2003 Kasner and Mixmaster behaviour in universes with equation
of state $w\ge 1$. Preprint hep-th/0312009.
\bibitem{foster} Foster, S. 1998 Scalar field cosmologies and the initial
spacetime singularity. Class. Quantum Grav. 15, 3485-3504.
\bibitem{kitada92} Kitada, Y. and Maeda, K. 1992 Cosmic no-hair theorem in
power-law inflation. Phys. Rev. D45, 1416-1419.
\bibitem{kitada93} Kitada, Y. and Maeda, K. 1993 Cosmic no-hair theorem in 
homogeneous spacetimes I. Bianchi models. Class. Quantum Grav. 10, 
703-734.
\bibitem{lee03} Lee, H. 2003 Asymptotic behaviour of the Einstein-Vlasov 
system with a positive cosmological constant. Preprint gr-qc/0308035.
\bibitem{lee04} Lee, H. 2004 Unpublished.
\bibitem{lin} Lin, X. and Wald, R. M. 1990 Proof of the closed-universe 
recollapse conjecture for general Bianchi type-IX cosmologies. Phys. Rev. D
41, 2444-2448.
\bibitem{miritzis} Miritzis, J. 2003 Scalar-field cosmologies with an 
arbitrary potential. Class. Quantum Grav. 20, 2981-2991.
\bibitem{moss} Moss, I. and Sahni, V. 1986 Anisotropy in the chaotic 
inflationary universe. Phys. Lett. B 178, 159-162.
\bibitem{mueller} M\"uller, V., Schmidt, H.-J. and Starobinsky, A. A.
1990 Power-law inflation as an attractor solution for inhomogeneous
cosmological models. Class. Quantum Grav. 7, 1163-1168.
\bibitem{peacock} Peacock, J. A. 1999 Cosmological physics. Cambridge 
University Press, Cambridge.
\bibitem{rendall02} Rendall, A. D. 2002 The Einstein-Vlasov system.
Preprint  gr-qc/0208082.
\bibitem{rendall03a} Rendall, A. D.  2003 Asymptotics of solutions of the 
Einstein equations with positive cosmological constant. Preprint 
gr-qc/0312020.
\bibitem{ringstrom2} Ringstr\"om, H. 2003 Future asymptotic expansions of 
Bianchi VIII vacuum metrics. Class. Quantum Grav. 20, 1943-1990.
\bibitem{ringstrom} Ringstr\"om, H. 2004 On a wave map equation arising in 
general relativity. Commun. Pure Appl. Math. (to appear).
\bibitem{starobinsky} Starobinsky, A. A. 1983 Isotropization of arbitrary
cosmological expansion given an effective cosmological constant. JETP Lett.
37, 66-69.
\bibitem{tchapnda03a} Tchapnda, B. and Noutchegueme, N. 2003 The 
surface-symmetric Einstein-Vlasov system with cosmological constant. Preprint 
gr-qc/0304098.
\bibitem{tchapnda03b} Tchapnda, B. and Rendall, A. D. 2003 Global existence 
and asymptotic behaviour in the future for the Einstein-Vlasov system with 
positive cosmological constant. Class. Quantum Grav. 20, 3037-3049.
\bibitem{wald} Wald, R. 1983 Asymptotic behaviour of homogeneous cosmological 
models with cosmological constant. Phys. Rev. D 28, 2118-2120.

\end{thebibliography}
\end{document}